\documentclass[12pt]{article}
\usepackage{amsmath, amssymb, amsthm}
\usepackage{graphicx}
\usepackage{enumerate}
\usepackage{natbib}
\usepackage{url} 
\usepackage[utf8]{inputenc}
\usepackage{multirow}
\usepackage{algorithm}
\usepackage{algorithmic}
\usepackage{float}
\usepackage{caption}
\usepackage{subcaption}
\usepackage{booktabs}
\usepackage{mathtools}
\usepackage{xcolor}
\usepackage{comment}

\newcommand{\blind}{1}

\newtheorem{definition}{Definition}
\newtheorem{proposition}{Proposition}

\begin{document}

\def\spacingset#1{\renewcommand{\baselinestretch}%
{#1}\small\normalsize} \spacingset{1}

\newcommand{\highlight}[1]{\textcolor{black}{#1}}
\newenvironment{revision}{\par\color{black}}{\par}

\if1\blind
{
  \title{\bf 
  Sparse Bayesian Modeling of EEG Channel Interactions Improves P300 Brain-Computer Interface Performance
  }
  \author{Guoxuan Ma\thanks{Equally contributed}, Yuan Zhong\footnotemark[1], Moyan Li\footnotemark[1],\  Yuxiao Nie, and Jian Kang\thanks{To whom correspondence should be addressed: jiankang@umich.edu}\\ Department of Biostatistics, University of Michigan, Ann Arbor MI 48109}  
  \maketitle
} \fi

\if0\blind
{
  \bigskip
  \bigskip
  \bigskip
  \begin{center}
    \LARGE\bf Sparse Bayesian Modeling of EEG Channel Interactions Improves P300 Brain-Computer Interface Performance
\end{center}
  \medskip
} \fi

\bigskip
\begin{abstract}
Electroencephalography (EEG)-based P300 brain-computer interfaces (BCIs) enable communication without physical movement by detecting stimulus-evoked neural responses. Accurate and efficient decoding remains challenging due to high dimensionality, temporal dependence, and complex interactions across EEG channels. Existing approaches often treat channels independently or rely on black-box models, limiting interpretability and personalization. We propose a sparse Bayesian time-varying regression framework that explicitly models pairwise EEG channel interactions while performing temporal feature selection, where a relaxed-thresholded Gaussian process prior induces structured sparsity in both channel-specific and interaction effects, enabling interpretable identification of task-relevant channels and channel pairs. Applied to a public P300 speller dataset of 55 participants, \highlight{our method achieves a 96.4\% median character-level accuracy using all sequence repetitions} and attains the best overall decoding performance among competing statistical and deep learning approaches. \highlight{Incorporating channel interactions yields subgroup-specific gains in character-level accuracy, particularly among participants abstained from alcohol (median 7\%, up to 14\%).} Importantly, our approach improves median BCI-Utility by more than 10\% at its optimal operating point, achieving peak throughput \highlight{after only six sequence repetitions}. These results demonstrate that explicitly modeling structured EEG channel interactions within a principled Bayesian framework enhances accuracy and user-centric throughput, and supports personalization in P300 BCI systems.
\end{abstract}

\noindent%
\vfill

\spacingset{1.8}

\section{Introduction}

Brain-computer interfaces (BCIs) facilitate direct communication between the human brain and external devices such as computers. The widespread interest in BCI systems stems from their broad range of potential applications in movement and communication assistance, particularly for individuals with motor impairments \citep{pfurtscheller2008rehabilitation}, while their applications include performance enhancement in healthy users, neurorehabilitation, and other cognitive and clinical research domains \citep{van2012brain,kwak2015lower}.

The BCI systems typically acquire brain activity through noninvasive electroencephalography (EEG). Among various EEG signals, event-related potentials (ERPs) refer to brain responses elicited by external stimuli such as visual, auditory, or somatosensory cues \citep{farwell1988talking}. In an ERP-based BCI design,  the system presents multiple stimuli in an on-screen keyboard, and the user focuses attention and mentally response to the desired stimulus. The stimulus that the user intends to select is called the target stimulus. After each stimulus presentation, the BCI classifies the corresponding EEG response as either target or non-target, depending on whether it contains ERP components of target perception. This design is often named after the P300 component, a deflection in the EEG signal that peaks approximately 300 ms after the onset of a rare or unexpected target stimulus \citep{fazel2012p300}. Among various applications of P300-based BCIs, the P300 speller is one of the most well-known. It functions as a virtual keyboard which allows users to type characters by attending to flashing groups of letters or symbols, which are then decoded based on the elicited P300 responses \citep{farwell1988talking}. \highlight{Section \ref{sec:data} provides a detailed description of our motivating dataset from \citet{won2022eeg}. }

The most-studied and fundamental computational challenge in P300-based BCIs is the classification of brain activity following each stimulus as either a target or non-target response. Accurate classification of these EEG segments enables the identification of the target row and column in the speller matrix, and hence the selection of the desired character. The original work \citep{farwell1988talking} developed four classification methods, stepwise linear discriminant analysis (SWLDA), peak picking, area, and covariance, with the best performance achieved by SWLDA in their experiment. Subsequent studies have sought to improve P300 speller performance through more advanced machine learning and signal processing techniques, including independent component analysis (ICA) \citep{xu2004bci} and support vector machines (SVMs) \citep{kaper2004bci}. A comprehensive review by \citet{philip2020visual} 
highlighted the strength of ensemble methods, which combine the advantages of multiple classifiers and are particularly effective for handling the class imbalance inherent in P300 datasets. Bayesian methods have also gained increasing attention for BCI classification. For example, \citet{zhang2015sparse} proposed a sparse Bayesian model using Laplace priors for EEG signal classification. \citet{barthelemy2023end} introduced a Bayesian accumulation of Riemannian probabilities, providing an end-to-end framework for P300 BCI classification. Recently, \citet{ma2022bayesian} developed a Bayesian generative model that characterizes the probability distribution of multi-trial EEG signals, which provides both a flexible simulation tool for EEG data and a novel probabilistic classifier for P300 BCIs.

Despite the success of existing methods, most approaches overlook potential functional relationships among brain regions and treat EEG signals from different channels as independent predictors. However, studies have shown that brain functions arise from the coordinated activity of distributed areas rather than isolated regions \citep{tononi1998consciousness,friston1997psychophysiological}. In the BCI context, \citet{kabbara2016functional} demonstrated clear differences in functional brain networks between target and non-target visual stimuli. These findings suggest that interactions among EEG channels, which reflect network-level brain dynamics, carry important information for distinguishing stimulus types (target/non-target). Modeling such inter-channel dependencies can therefore enhance both the predictive accuracy and the neurophysiological interpretability of BCI systems.

To model signal interactions, many existing approaches adopt a two-step strategy, by first identifying main effects and then refitting the model with both main and interaction effects \citep{hao2018model,wang2021penalized}. However, this approach is less suitable for EEG data, as the presence of an interaction does not necessarily depend on the existence of corresponding main effects. \citet{zhao2025bayesian} took a difference approach by introducing the Gaussian Latent channel model with Sparse time-varying effects (GLASS) for P300 speller, which uses constrained multinomial logistic regression for target classification while accounting for correlations between channels via latent channel decompositions. Although GLASS incorporates channel correlations into the model structure, it does not explicitly model channel interactions as predictors, and therefore cannot directly evaluate and interpret the effect sizes of inter-channel relationships. Recent Bayesian approaches have incorporated both main and interaction terms within a unified inference framework using hierarchical shrinkage priors \citep{griffin2017hierarchical}. Nonetheless, such general frameworks is less effective in accounting for the unique temporal and structural dependencies in EEG data. More recently, neural network-based methods have been proposed in EEG classification tasks. Examples include multi-task autoencoder models \citep{ditthapron2019universal}, compact convolutional neural networks such as EEGNet \citep{lawhern2018eegnet}, and weighted ensemble strategies \citep{kshirsagar2019weighted}. These approaches can implicitly capture interaction effects among EEG channels and their association with stimulus type outcomes. However, they often require extensive task-specific architectural design and large training datasets, which vary considerably across studies. Moreover, the black-box nature of neural networks limits interpretability, making it difficult to identify which specific channels or channel pairs drive classification decisions. This lack of interpretability constrains the scientific insights that can be drawn about the underlying neural mechanisms and may hinder model generalizability. Consequently, there is a strong need for statistical methods that can incorporate inter-channel interaction effects while preserving interpretability and computational efficiency.

In this paper, we propose a Bayesian time-varying regression model with signal interactions via relaxed-thresholded Gaussian process (SIRTGP) priors. Our model relaxes the traditional linearity assumption among EEG predictors by explicitly modeling signal interaction effects across channels, while performing temporal-spatial channel selection, thereby improving both predictive performance and interpretability. To our knowledge, this is among the first models to explicitly incorporate inter-channel interactions into EEG-based prediction within a Bayesian framework. 

We propose a relaxed-thresholded Gaussian process (RTGP) prior to flexibly model the association between EEG signals and stimulus-type outcomes. It has several advantages compared to existing work. First, the RTGP prior defines a broad class of temporally varying functions that are piecewise smooth and sparse, which enables automatic feature selection with Bayesian inference. Second, compared with existing thresholded Gaussian process priors, such as the soft-thresholded \citep{kang2018scalar} and hard-thresholded \citep{cai2020bayesian} variants, the proposed RTGP prior is more flexible, capable of adapting to both sparse and non-sparse patterns by tuning a relaxation parameter. It also offers substantial computational advantages, allowing efficient MCMC sampling even for large-scale EEG datasets. We evaluate the proposed method using both synthetic data and EEG data from the P300 speller study by \citet{won2022eeg}. The SIRTGP model achieves higher classification accuracy and throughput metrics, and identifies meaningful channels and channel pairs, which provides valuable insights into the neural mechanisms underlying P300 responses.

\begin{revision}
\section{Dataset and Motivation}\label{sec:data}
\subsection{Dataset}

In this section, we describe the motivating EEG dataset from a publicly available P300-based BCI experiment by \citet{won2022eeg}. The study collected EEG recordings from 55 participants during BCI spelling tasks, together with questionnaire data on demographics and physical and mental states before and after the experiment. Each participant completed a training phase (two runs) and a testing phase (four runs) using a P300 speller. In the two training runs, participants spelled the words “BRAIN” and “POWER”. In the testing runs, they spelled four additional words, “SUBJECT,” “NEURONS,” “IMAGINE,” and “QUALITY”. 

During spelling, rows and columns of a virtual keyboard (a 6$\times$6 matrix; see Figure \ref{fig:bci}A) were flashed repeatedly in a random order. We call a round of the 12 flashes/stimuli (i.e., 6 rows and 6 columns) a sequence. Each character selection included 15 such sequence repetitions in both training and testing runs. For a given target character, the flashes corresponding to its row and column constituted target stimuli, while all remaining flashes were non-target.  Each stimulus was displayed for 125 ms, followed by a 62.5 ms pause. 

\begin{figure}[!t]
    \spacingset{1}
    \small
    \centering
    \includegraphics[width=0.9\linewidth]{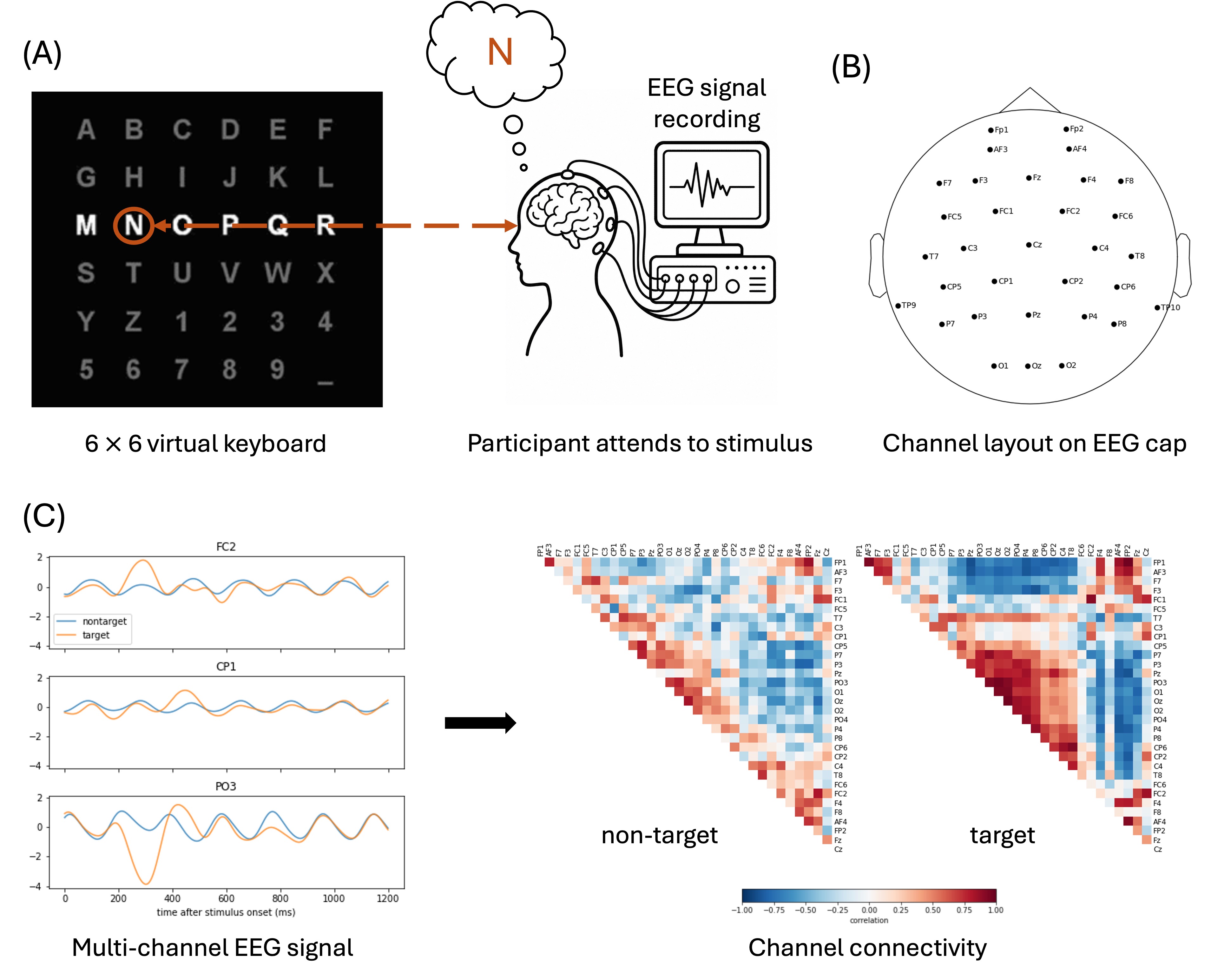}
    \caption{\highlight{An illustration of the P300 BCI spelling task and recorded signal. (A) The P300 BCI speller presents a sequence of row or column flashes (a row or a column) on a virtual screen to the user. The user focuses on a target character and responds to the row and column flashes that contain the target character. The EEG signals of the flash are recorded from multiple channels and are segmented into a 1200 ms window following stimulus onset. (B) The names and layout of the 32 scalp EEG channels according to the international 10-20 system. (C) Example of recorded multi-channel EEG signals and the derived channel connectivity.}}
    \label{fig:bci}
\end{figure}

EEG signals were recorded from 32 scalp channels arranged according to the international 10-20 system, sampled at 512 Hz, and segmented into 1200 ms windows following the stimulus onset. EEG preprocessing included a bandpass filter with a frequency range of 0.5-10 Hz. Although the dataset provides a 1200 ms post-stimulus window, we restrict our analysis to the first 600 ms after stimulus onset, corresponding to 307 samples at the 512 Hz sampling rate, which is sufficient to capture the event-related responses associated with the P300 component. As a result, for each participant, the EEG signal in training set has dimensions of $32\times307\times300$ for target stimuli, corresponding to 32 channels, 307 time points per flash, and 300 target flashes (10 training letters$\times$15 sequences$\times$2 target flashes per sequence), and $32\times307\times1500$ for non-target. The EEG signal in testing set has dimensions of $32\times307\times840$ for target stimuli and $32\times307\times4200$ for non-target. The response variable is a binary vector of whether or not each flash is a target stimulus, with length 1800 (10 training letters$\times$15 sequences$\times$12 flashes per sequence) in training set and 5040 in testing set, for each participant. 

\subsection{Motivations and challenges}

Figure \ref{fig:bci}C shows the EEG recording from three example channels (left) and the derived channel connectivity from the multi-channel EEG signal recording (right). As illustrated in the left panel of Figure \ref{fig:bci}C, discriminative information between the signals recorded after target and non-target stimulus varies across channels and time. For example, channels FC2 and PO3 show clear target-related deflections in the P300 latency range, whereas channel CP1 exhibits much weaker differences. Even within informative channels, only limited post-stimulus time intervals contribute meaningfully to discrimination. These motivate temporal channel selection via sparse modeling to reduce noise from uninformative channels and time points. 

Beyond marginal temporal features, the signal interactions between EEG channels, summarized through measures of channel connectivity, provide an additional and complementary source of discriminative information. The right panel of Figure \ref{fig:bci}C shows that the connectivity patterns derived from the EEG signal associated with the target stimulus differ from those of non-target. Incorporating channel connectivity as an additional predictor, alongside temporal EEG features, therefore has the potential to improve classification performance. 

Furthermore, because subject-level questionnaire data on demographics and physical and mental states are available, this dataset enables investigation of heterogeneity across individuals, including whether certain subgroups of participants benefit more from accounting for signal interactions in modeling than others. This opens the door to identifying populations for whom modeling channel interactions is particularly advantageous, with implications for personalized and adaptive BCI system design.
\end{revision}

\section{Method}

\subsection{Bayesian time-varying model with signal interactions}

Our model is individual-specific, meaning that a separate model is built for each participant using only their own calibration data and evaluated on their corresponding test data. For clarity, the subject index is suppressed in the notation throughout the model development. 

\begin{figure}[t!]
\spacingset{1}
\centering
\includegraphics[width=\textwidth]{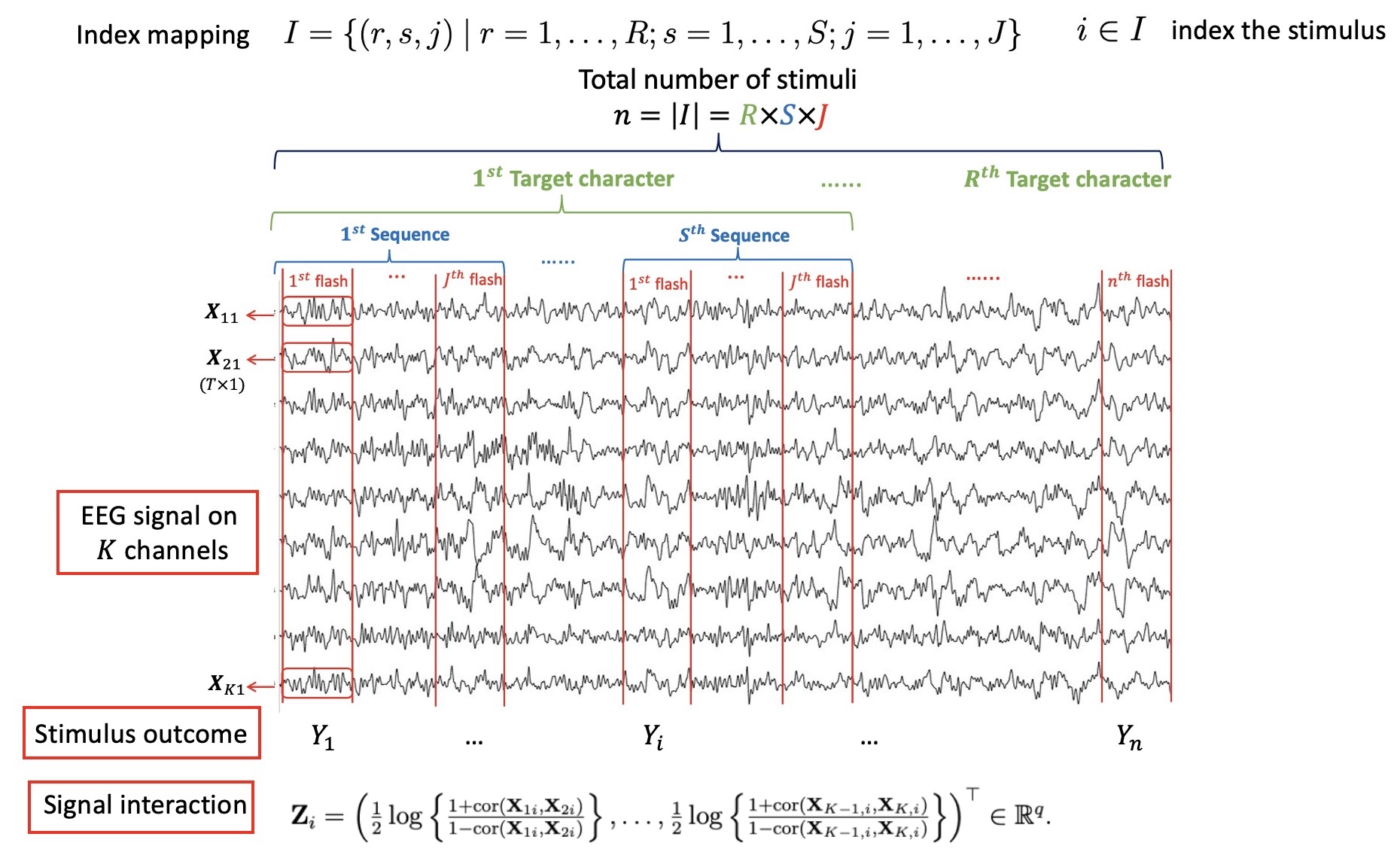}
\caption{Illustration of the data and notation for the SIRTGP model. $\mathbf{X}_{ki}$ represents the EEG signal from channel $k$ for the $i$-th flash. For the $i$-th flash, the signal interaction between channel $k_1$ and channel $k_2$ is measured by the z-transformed Pearson correlation between EEG signals from channels $k_1$ and $k_2$.}
\label{fig::data_pre}
\end{figure}

Suppose a total of $R$ target characters are typed during the calibration phase. For each character $r=1,\ldots,R$, the BCI presents $S$ \highlight{sequence repetitions}, each consisting of $J=12$ flashes: 6 row stimuli ($j=1, \ldots, 6$) and 6 column stimuli ($j=7, \ldots, 12$) on a $6 \times 6 $ speller matrix, in a random order. Let $I = \{(r, s, j)\ |\ r = 1, \ldots, R; s = 1, \ldots, S; j = 1, \ldots, J\}$ denote the index set of all stimulus presentations. The total number of flashes is therefore $n = |I| = R\times S\times J$ flashes in total, and we use $i\in I$ to index an individual flash. Let $K$ denote the number of EEG channels and $T$ the number of time points within the post-stimulus window. Then, denote $X_{ki}(t)$ the observed intensity of the EEG signal of the $i$-th stimulus from channel $k$ at time $t$. Let $\mathbf{X}_{ki} = (X_{ki}(1), \ldots, X_{ki}(T))^\top \in \mathbb{R}^T$ and $\mathbf{X}_i = \left(\mathbf{X}_{1i}^\top, \mathbf{X}_{2i}^\top,\cdots, \mathbf{X}_{Ki}^\top\right)^\top \in \mathbb{R}^{KT}$. We denote by $Z_{i}(k_1, k_2)$ the signal interaction between channels $k_1$ and $k_2$, where $1 \leq k_1< k_2 \leq K$. In this study, we define the signal interaction as the Fisher Z-transformation of the Pearson correlation between $\mathbf{X}_{k_1i}$ and $\mathbf{X}_{k_2i}$, i.e. 
$$Z_{i}(k_1, k_2) = \frac{1}{2}\log\left\{\frac{1 + \mathrm{cor}(\mathbf{X}_{k_1i}, \mathbf{X}_{k_2i})}{1 - \mathrm{cor}(\mathbf{X}_{k_1i}, \mathbf{X}_{k_2i})}\right\}.$$ 
Let $\mathbf{Z}_{i} = \{Z_{i}(k_1, k_2)\}_{1 \leq k_1< k_2 \leq K}$ be a vector of length $K(K-1)/2$. Finally, let $Y_i \in\{0,1\}$ be the target/non-target stimulus type outcomes of the $i$-th flash. Figure \ref{fig::data_pre} provides an illustration on data and notations. 


We propose a Bayesian time-varying classification model with signal interactions to predict the stimulus type outcome, as follows,
\highlight{
\begin{equation}\label{eq: main_model}
        \mathrm{g}\left\{\operatorname{Pr}(Y_i=1  \mid \mathbf{X}_i, \mathbf{Z}_{i})\right\} =\dfrac{1}{KT}\sum_{k=1}^K\sum_{t=1}^{T}\beta_k(t)X_{ki}(t) + \dfrac{1}{K(K-1)/2}\sum_{k_1<k_2}\zeta(k_1, k_2)Z_{i}(k_1, k_2), 
\end{equation}
}
where $\mathrm{g}(\cdot)$ is the link function (e.g., probit and logit), $\beta_k(t)$ is a time-varying coefficient function for channel $k$, and $\zeta(k_1, k_2)$ quantifies the effect of the signal interaction between the EEG signal from channels $k_1$ and $k_2$. The rescaling factors \highlight{$KT$ and $K(K-1)/2$} are introduced to prevent the linear predictor $\mu_i$ from becoming excessively large when $K$ and $T$ are large. Model \eqref{eq: main_model} accounts for signals from each separate channel and signal interactions across channels as predictors for stimulus type outcomes classification. 

\subsection{Relaxed-thresholded Gaussian process prior}

Gaussian process priors are useful in modeling and capturing dependencies in functional data, and they have been applied successfully in numerous neuroimaging studies \citep{lin2024latent,wu2025bayesian,ma2026bayesian,he2026scalable}. Among their variants, thresholded Gaussian process priors are particularly effective at capturing both the sparsity and temporal dependency inherent in the relationship between stimulus type and EEG signal predictors. However, existing approaches, including the soft-thresholded Gaussian process \citep{kang2018scalar,xubayesian,xu2026bayesian} and the thresholded Gaussian process \citep{shi2015thresholded,wu2024bayesian}, pose significant computational challenges when applied to large-scale datasets. To address this limitation, we propose a relaxed-thresholded Gaussian process (RTGP) prior, which preserves the ability to capture sparsity while substantially improving computational efficiency. The RTGP prior is defined as follows.

\begin{definition}\label{definition: RTGP}
Given a kernel $\kappa$, a thresholding parameter $\omega \geq 0$, and a relaxing parameter $\xi>0$, suppose $f(x)\sim \textrm{GP}(0,\kappa)$ and $\Tilde{f}(x)\sim \textrm{N}(f(x),\xi^2)$. Let $g(x) = f(x)I(|\Tilde{f}(x)| > \omega) \triangleq T_r(f, \omega, \xi^2)$, then $g(x)$ follows a relaxed-thresholded Gaussian process, denoted as $g(x)\sim \textrm{RTGP}(\kappa, \omega, \xi^2)$.
\end{definition}

In Definition \ref{definition: RTGP}, $I(\cdot)$ denotes the indicator function and $T_r$ is the relaxed-thresholding function. The introduction of $\tilde{f}(x)$ allows the full conditional distribution of $f(x)$ to have a conjugate and closed-form expression. \highlight{Although RTGP is constructed from an underlying Gaussian process, it is generally not a GP. In particular, when the thresholding parameter $\omega=0$, RTGP becomes the original GP as a degenerate case. For RTGP with $\omega>0$, the thresholding operation yields a non-Gaussian process that remains a well-defined prior over the functional space while enabling local sparsity.}

\begin{figure}[t!]
\spacingset{1}
\centering
\includegraphics[width=0.7\textwidth]{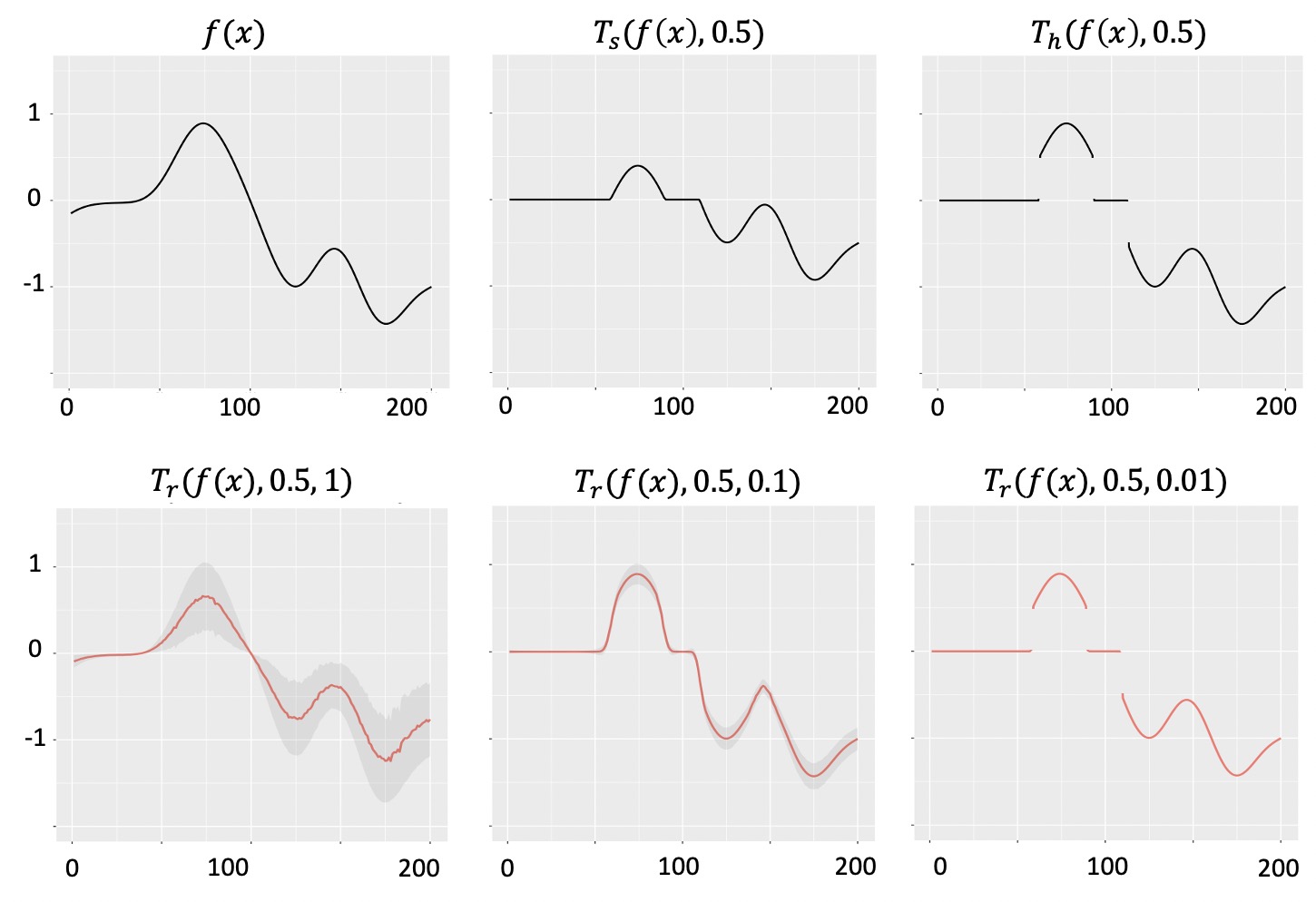}
\caption{\highlight{Illustration of different thresholded Gaussian process prior. The top-left panel shows a draw $f(x)$ from a Gaussian process prior. The remaining panels show the effect of three thresholding operators with threshold parameter $\omega=0.5$: soft thresholding $T_s$, hard thresholding $T_h$, and the proposed relaxed thresholding $T_r$. The bottom row illustrates the relaxed thresholding $T_r$ for decreasing relaxation parameters $\xi^2\in\{1, 0.1, 0.01\}$. Shaded bands show 95\% credible intervals; red lines are the sample means.}}
\label{fig::prior}
\end{figure}

The parameter $\xi^2$ represents the variance of $\tilde{f}(x)$ and serves as a relaxing parameter that controls the independent white noise added to ${f}(x)$. Smaller values of $\xi^2$ impose a stricter constraint that preserves the mean structure of ${f}(x)$, while larger values provide greater flexibility. To illustrate, Figure \ref{fig::prior} compares different thresholded GP functions. Let $f(x)\sim \mathrm{GP}(0, \kappa)$. The soft-thresholding function $T_s(f(x), 0.5)$ sets values with $|f(x)| < 0.5$ to zero, and otherwise shrinks the magnitude by the threshold 0.5. The hard-thresholding function $T_h(f(x), 0.5)$ also sets values below magnitude of 0.5 to zero, while leaving larger values unchanged. Both soft and hard thresholded GPs impose sparsity and piecewise smoothness, with the hard-thresholded GP introducing jump discontinuities and the soft-thresholded GP remaining continuous. The second row of Figure \ref{fig::prior} illustrates the proposed RTGP under different values of $\xi$. 
When $\xi = 0.01$, $T_r\{f(x), 0.5, 0.01\}$ closely resembles the hard-thresholded GP. For $\xi = 0.1$, $T_r\{f(x), 0.5, 0.1\}$ is a continuous function that preserves sparsity, similar to the soft-thresholded GP. As $\xi$ increases to 1, $T_r\{f(x), 0.5, 1\}$ recovers $f(x)$. 

This flexibility is particularly valuable in modeling EEG data, where the true signal pattern is unknown. For example, if the true curve near $x=100$ is with meaningful magnitude as in \ref{fig::prior}, RTGP can adapt and recover the pattern by choosing an appropriate $\xi$ (e.g., $\xi=1$), whereas both soft- and hard-thresholded GPs would enforce sparsity with probability 1. The following proposition formalizes the relationship between the RTGP and other thresholded GPs.

\begin{proposition}\label{prop:RTGP}
Given a thresholding parameter $\omega>0$, let $T_r(\theta, \omega, \xi^2) = \theta\cdot I(|\Tilde{\theta}|>\omega)$, $T_h(\theta, \omega) = \theta\cdot I(|{\theta}|>\omega)$ and $T_s(\theta, \omega) = \operatorname{sgn}(\theta)(|\theta| - \omega)\cdot I(|{\theta}|>\omega)$ where $\theta \sim \mathcal{P}_\theta(\theta)$. Then for any $\epsilon > 0$, there exist $\xi^2$, such that     $\operatorname{Pr}\left(|T_r(\theta, \omega, \xi^2) - T_h(\theta, \omega)| < \epsilon\right) > 0$,  $\operatorname{Pr}\left(|T_r(\theta, \omega, \xi^2) - \theta| < \epsilon\right) > 0$, and $\operatorname{Pr}\left(|T_r(\theta, \omega, \xi^2) - T_s(\theta^\star, \omega)| < \epsilon\right) > 0$,
where $\theta^\star = \theta + \omega$ when $\theta > 0$ and $\theta^\star = \theta - \omega$ when $\theta < 0$.
Furthermore, $\lim\limits_{\xi^2 \to 0} T_r(\theta, \omega, \xi^2) = T_h(\theta, \omega)$ and $\lim\limits_{\xi^2 \to \infty} T_r(\theta, \omega, \xi^2) = \theta$.
\end{proposition}
Proposition \ref{prop:RTGP} provides a mathematical illustration of Figure \ref{fig::prior}, which shows that relaxed-thresholded function has certain probability to reduce to soft or hard thresholded function and the flexibility is controlled by the relaxing parameter $\xi^2$. The proof is included in Section S2 of Supplementary Materials. 

Given a stationary kernel $\kappa$, we assign $\beta_k(t) \sim \textrm{RTGP}(\kappa, \omega_1, \xi^2)$ for channel $k$. That is,  
\begin{equation}
\label{eq: prior_beta}
        \beta_k(t) = E_{k}(t) I(|\Tilde{E}_{k}(t)| > \omega_1),\quad        E_k(t) \sim \mathrm{GP}(0, \kappa),\quad        \tilde{E}_k(t)\sim \mathrm{N}(E_k(t), \xi^2),
\end{equation}
where $\{E_k(\cdot)\}_{k=1}^K$ follow independent Gaussian processes. \highlight{Throughout this paper, we employ the modified squared exponential (MSE) kernel for $\beta_k(\cdot)$ \citep{wu2024bayesian,wu2025bayesian}. Details of the MSE kernel and the associated computational strategies are deferred to Section \ref{sec:posterior_mse}.}

Similarly, for the effect of signal interaction across channels,  we assign $\zeta(k_1, k_2)\sim \textrm{RTGP}(\sigma^2_\eta\kappa_I ,\omega_2, \xi^2), k_1<k_2$, where $\kappa_I$ represents the identity kernel. That is, 
\begin{equation}
\label{eq: prior_beta0}
    \begin{split}
        &\zeta(k_1, k_2) = \eta(k_1, k_2) I(|\Tilde{\eta}(k_1, k_2)|> \omega_2), \\ 
        &\eta(k_1, k_2) \stackrel{\mathrm{iid}}{\sim} \mathrm{N}(0, \sigma_\eta^2),\qquad
        \Tilde{\eta}(k_1, k_2) \sim \mathrm{N}(\eta(k_1, k_2), \xi^2),
    \end{split}  
\end{equation}
and $\sigma_\eta^2 \sim \mathrm{IG}(a_\eta,b_\eta)$. We choose to use the identity kernel here to assume prior independence across channel pairs. 
Other kernels can be adopted when prior information is available in different applications. 

By combining Model \eqref{eq: main_model} with the prior specifications in \eqref{eq: prior_beta} and \eqref{eq: prior_beta0}, we define the proposed model as a Bayesian time-varying classification model with signal interactions via relaxed-thresholded Gaussian process prior (SIRTGP). 

\section{Posterior Computation}
\label{sec:posterior}
In this section, we describe posterior computations for Model \eqref{eq: main_model} with priors \eqref{eq: prior_beta} and \eqref{eq: prior_beta0}. 

\subsection{Equivalent model representation}

We represent the Gaussian processes by the Karhunen-Lo\`eve expansion and obtain an equivalent model representation. Specifically, consider the spectral decomposition of the kernel function,  $\kappa\left(x, x^{\prime}\right)=\sum_{l=1}^{\infty} \lambda_l \psi_l(x) \psi_l\left(x^{\prime}\right)$, 
where $\{\lambda_{l}\}_{l=1}^{\infty}$ are the eigenvalues in descending order, and $\{\psi_{l}(x)\}_{l=1}^{\infty}$ are the corresponding orthonormal eigenfunctions. 
By Mercer’s Theorem, we can represent the Gaussian process $E_k(t)$ in \eqref{eq: prior_beta} by $E_k(t) = \sum_{l=1}^\infty e_{kl}\psi_l(t)$, 
where $e_{kl}$ are Karhunen-Lo{\`e}ve coefficients. We truncate the expansion to the leading $L$ terms, where $L$ is chosen following the common practice in principal component analysis where we retain enough components to explain a high proportion of total variation. Then, the prior of $\beta_k(\cdot)$ specified in \eqref{eq: prior_beta} becomes 
\begin{align}
    \beta_k(t) = \left\{\sum_{l=1}^L e_{kl}\psi_l(t)\right\} I(|\Tilde{E}_{k}(t)| > \omega_1), \quad
        \tilde{E}_k(t) \sim \mathrm{N}\left(\sum_{l=1}^L e_{kl}\psi_l(t), \xi^2\right), \label{eq:equivalent_representation}
\end{align}
and $e_{kl} \sim \mathrm{N}(0,\sigma_e^2 \lambda_l)$. 

\subsection{The modified squared-exponential kernel}\label{sec:posterior_mse}
\highlight{We use the modified squared exponential (MSE) kernel \citep{wu2024bayesian,wu2025bayesian}, defined as 
$$\kappa\left(x, x^{\prime}\right)=\exp \left\{-\alpha\left(\lVert x\rVert_2^2+\left\lVert x'\right\rVert_2^2\right)-\rho\left\lVert x-x'\right\rVert_2^2\right\},$$
where $\alpha>0$ controls the decay rate of how fast the variance decays relative to the origin $x=0$ and $\rho >0$ controls the smoothness of the kerne. The MSE kernel reduces to a standard squared exponential kernel when $\alpha = 0$. A smaller $\alpha$ corresponds to a slower decay rate, while a smaller value of $\rho$ corresponds to a smoother kernel. We adopt the MSE kernel for both computational and practical reasons. Computationally, it admits a closed-form basis decomposition, with basis functions $\psi_l$ represented by Hermite polynomials. Supplementary Materials Section S3 provides details. This closed-form representation is not generally available for many alternative kernel families and provides a substantial computational advantage in high-dimensional functional modeling. Practically, the MSE kernel has shown considerable flexibility in modeling functional effects and has been successfully applied in neuroimaging and BCI studies \citep{lin2024latent,wu2025bayesian,lin2026spatial}. Following these works, we set $\alpha=0.01$. We estimate $\rho$ following the discussion in \citet{lin2024latent}. Specifically, $\rho$ is estimated by averaging the smoothing parameters obtained from fitting GP models to the EEG signals across all channels. We also include a sensitivity analysis with respect to $\rho$ in the Supplementary Materials Section S5.}

\subsection{Prior specification}

We set \highlight{the variance of basis coefficients} $\sigma_e^2$ to be large values, $a_\eta=b_\eta=0.001$ \highlight{in the Inverse Gamma prior of $\sigma^2_\eta$ so that it is non-informative}. For flexibility, we set \highlight{the relaxing paramter} $\xi^2 = 1$ in the first 200 iterations and then gradually decrease its value to $0.0001$. We set \highlight{thresholding parameters} $\omega_1$ and $\omega_2$ to zero in the first 200 steps, then assign an adaptive discrete prior, i.e., $P(\omega_1 = \gamma_{1z}) = 1/Z$ and $P(\omega_2 = \gamma_{2z}) = 1/Z$, $z = 1, \cdots, Z$, where $\{\gamma_{1z}\}_{z=1}^Z$ and $\{\gamma_{2z}\}_{z=1}^Z$ are $Z$ evenly spaced number between $a_\omega$ quantile and $b_\omega$ quantile of $\{|\Tilde{E}_k(t)|\}_{k=1, t=1}^{K,T}$ and $\{|\Tilde{\eta}(k_1, k_2)|\}_{k_1<k_2}$ respectively. In practice, we found $a_\omega = 0.25$ and $b_\omega = 0.90$ yield good performance. We include the sensitivity analysis of $a_\omega$ and $b_\omega$ in the Supplementary Materials Section S5. 

Given the above specifications, the full conditionals of $\{e_{k,l}\}_{k=1, l=1}^{K,L}$, $\{\Tilde{E}_k(t)\}_{t=1,k=1}^{T,K}$, $\{\eta(k_1, k_2)\}_{k_1<k_2}$ and $\{\Tilde{\eta}(k_1, k_2)\}_{k_1<k_2}$ are normal distributions due to conjugacy. The full conditionals of $\omega_1$ and $\omega_2$ are discrete distribution. We develop a Gibbs sampler for the posterior sampling. The detailed derivation and the sampling procedure is provided in Supplementary Materials S1. 

\highlight{We evaluate the proposed method through extensive simulation studies under a variety of settings characterized by different signal magnitudes, noise variances, and interaction strengths. The proposed approach is compared with competing statistical, machine learning, and deep learning methods for EEG-based BCI task. Results demonstrate that our method consistently achieves superior predictive and selection performance. Additional details of the simulation design and results are provided in Section S4 in the Supplementary Materials.}

\section{P300 BCI Speller Data Analysis}\label{sec::real}

We analyze a publicly available P300 BCI dataset consisting of 55 participants \citep{won2022eeg}. \highlight{We first introduce competing methods and their implementation details, and then compare the prediction performance across all methods. Using accompanying questionnaire data, we then examine which population subgroups benefit more from modeling signal interactions and discuss heterogeneity in channel interaction effects. Finally, we illustrate how the proposed method performs dynamic temporal channel selection by identifying informative channels over time.}

\subsection{Benchmark methods and implementation details}

We compare four model variants based on the proposed method, RTGP-L, RTGP-P, SIRTGP-L, and SIRTGP-P, with several commonly used classification methods. The four variants are Bayesian time-varying regression approaches that incorporate relaxed-thresholded Gaussian Process priors, using either logit (-L) or probit (-P) link functions. RTGP models include only main effects from individual EEG channels, while SIRTGP models additionally incorporate pairwise signal interactions across channels. 

For benchmarking, we include GLASS \citep{zhao2025bayesian}, EEGNet \citep{lawhern2018eegnet}, extreme gradient boosting (XGBoost) \citep*{sarraf2023study}, logistic regression (LR) \citep{sakamoto2009supervised}, support vector classification (SVC) \citep{kaper2004bci}, random forest (RF), and stepwise linear discriminant analysis (SWLDA) \citep*{farwell1988talking}. The GLASS model is trained using the recommended default settings from the authors' GLASS implementation \citep{zhao2025bayesian}. LR is implemented using the scikit-learn LogisticRegression class with a maximum of 3000 iterations. SVC is implemented with the SVC class with probability estimation enabled and also uses a maximum of 3000 iterations. The random forest model uses 1000 trees, with a maximum tree depth of 3. At each split, the number of features considered is set to the square root of the total number of input features. The XGBoost model is configured with 1000 boosting rounds, a learning rate of 0.02, a maximum tree depth of 5, a column subsampling ratio of 0.8, and a regularization parameter $\gamma = 10$, using the binary logistic loss function. EEGNet is trained for 100 epochs with a batch size of 32, a dropout rate of 0.5, and a kernel length of 128. SWLDA is implemented using forward-backward selection with entry and removal $p$-value thresholds set to 0.05 and 0.1, respectively. \highlight{For the probit-link variants of our framework, we used 1,000 post-burn-in iterations (following 5,000 burn-in iterations) for estimation. For the logit-link variants, we used all 6,000 MCMC iterations for estimation. Convergence was assessed using the Gelman-Rubin diagnostic based on three chains per participant. Across all four model variants, $\hat{R}$ values indicated convergence ($\hat{R} < 1.1$) for at least 85\% of participants (47/55).}

\subsection{Evaluation of prediction performance}

\highlight{We train all methods using data from the training runs and evaluate the prediction performance with data from testing runs.} The stimulus-specific prediction output by each method, such as predicted probabilities or discrimination scores, is treated as a classification score that quantifies how likely each stimulus group (one row or one column) contains the target character. To evaluate character-level prediction accuracy, we aggregate these classification scores across sequences. Specifically, for each character, we sum the scores for all row and column stimuli across sequences and identify the predicted character as the one located at the intersection of the row and column with the highest cumulative score. To provide a more comprehensive assessment of model performance, we repeat this evaluation using the first 1 through 15 sequences. This allows us to examine how prediction accuracy improves as more sequences (i.e., repetitions of the flashing paradigm) become available.

\begin{figure}[t]
    \spacingset{1}
    \centering
    \includegraphics[width=0.8\linewidth]{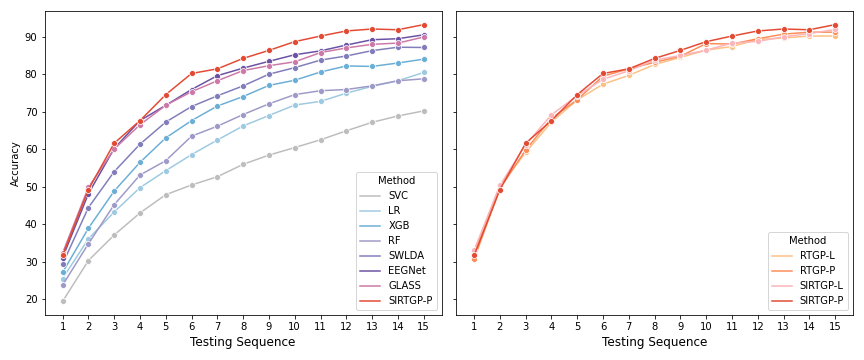}
    \caption{\highlight{Mean character prediction accuracy across 55 participants for different methods over 15 sequences using the proposed models (four variants) and competing methods. For clarity, the left panel includes the proposed method with channel interaction and probit link (SITRGP-P) and competing methods, while the right panel compares the four model variants. Competing methods include Support Vector Classifier (SVC), Logistic Regression (LR), Extreme Gradient Boosting (XGBoost), Random Forest (RF), Stepwise Linear Discriminant Analysis (SWLDA), EEGNet and GLASS}.}
    \label{fig:real_accuracy_55}
\end{figure}

\begin{table}[t]
    \spacingset{1}
    \small
    \centering
    \caption{Character prediction accuracy for all methods at the last (15th) sequence. \textbf{Mean}: mean accuracy; \textbf{SD}: standard deviation; \textbf{Median}: median accuracy; \textbf{Q1}: 1st quartile for accuracy; \textbf{Q3}: 3rd quartile for accuracy.}
    \begin{tabular}{l c c }
    \hline\hline
    \textbf{Method} & \textbf{Mean (SD)} & \textbf{Median (Q1, Q3)} \\
    \hline
    \textbf{EEGNet} & 90.58\% (13.65\%) & 96.43\% (89.29\%, 100.00\%) \\
    \textbf{GLASS} & 90.00\% (14.11\%) & 92.86\% (89.29\%, 100.00\%) \\
    \textbf{LR} & 80.52\% (20.51\%) & 85.71\% (71.43\%, 96.43\%) \\
    \textbf{RF} & 78.83\% (19.73\%) & 82.14\% (66.07\%, 94.64\%) \\
    \textbf{SVC} & 70.26\% (26.56\%) & 78.57\% (57.14\%, 91.07\%) \\
    \textbf{SWLDA} & 87.21\% (15.53\%) & 89.29\% (82.14\%, 96.43\%) \\
    \textbf{XGB} & 84.09\% (15.23\%) & 85.71\% (75.00\%, 96.43\%) \\
    \textbf{RTGP-L} & 90.26\% (13.12\%) &	96.43\%	(85.71\%, 100.00\%) \\
    \textbf{RTGP-P} & 91.30\% (13.37\%) & 96.43\% (89.29\%, 100.00\%) \\
    \textbf{SIRTGP-L} & 92.01\% (12.45\%) & 96.43\% (89.29\%, 100.00\%) \\
    \textbf{SIRTGP-P} & \textbf{93.31\% (11.00\%)} & \textbf{96.43\% (92.86\%, 100.00\%)} \\
    \hline\bottomrule
    \end{tabular}
    \label{tab:real_accuracy}
\end{table}

\highlight{Figure~\ref{fig:real_accuracy_55} presents the mean character prediction accuracy across 55 participants for different methods over 15 sequence repetitions.} Among the competing approaches, EEGNet and GLASS achieve the highest accuracies. However, SIRTGP-P has the highest mean accuracy at every sequence comparing to competing methods. The four model variants based on the proposed method have comparable performance, with the variants using the probit link outperforming those using the logit link. In addition, the model variants using signal interactions achieve equal or higher mean accuracy than those not using interactions at most sequences. Table \ref{tab:real_accuracy} summarizes mean, standard deviation, median, 1st and 3rd quartiles of the character prediction accuracy using all the 15 sequences for all methods. Overall, SIRTGP-P achieves the best prediction performance, with the highest mean and median accuracy and with relatively small variations across participants. All four model variants of the proposed method outperform competing methods, with equal or higher mean and median prediction accuracy and comparatively small performance variation. 

\begin{figure}[t]
    \spacingset{1}
    \centering
    \includegraphics[width=0.8\linewidth]{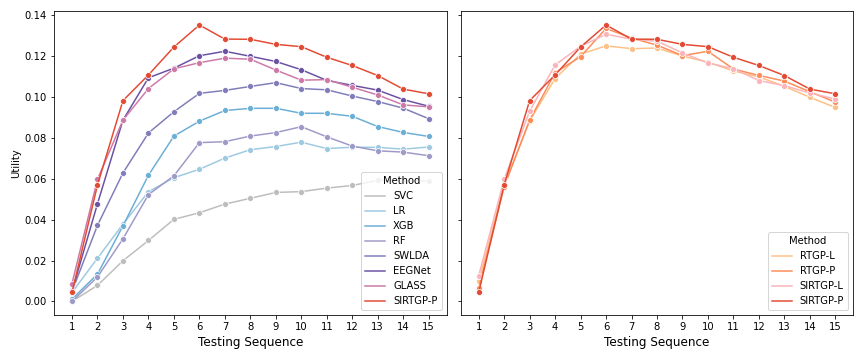}
    \caption{\highlight{Mean BCI-Utility (bit/second) across 55 participants for different methods over 15 sequences using the proposed models (four variants) and competing methods. For clarity, the left panel includes the proposed method with channel interaction and probit link (SITRGP-P) and competing methods, while the right panel compares the four model variants. Competing methods include Support Vector Classifier (SVC), Logistic Regression (LR), Extreme Gradient Boosting (XGBoost), Random Forest (RF), Stepwise Linear Discriminant Analysis (SWLDA), EEGNet and GLASS}. }
    \label{fig:real_utility_55}
\end{figure}

Although character-level accuracy is a fundamental performance metric for BCI systems, it does not fully reflect the user experience. In practice, users care not only about achieving high accuracy but also about how quickly that an accurate selection can be made \citep{ma2023bci}. To complement the accuracy results in Figure~\ref{fig:real_accuracy_55}, we additionally evaluate system throughput using the BCI-Utility metric. BCI-Utility is a user-centric measure that quantifies the average information gained per unit time and is maximized when a system provides both high accuracy and rapid decision-making \citep{dal2009utility, ma2023bci}. We compute the BCI-Utility (bit/second) for all methods using available signal up to each of the 15 sequences.

\highlight{Figure~\ref{fig:real_utility_55} shows the mean BCI-Utility across 55 participants.} Consistent with the accuracy results, the proposed SIRTGP-P with signal interaction achieves the highest mean BCI-Utility at every sequence and provides the best overall throughput among all competing methods and model variants. Importantly, BCI-Utility and accuracy do not increase monotonically over sequence in the same way. While accuracy continues to improve as more sequences of data becomes available, BCI-Utility typically peaks earlier because additional sequences increase decision time. For SIRTGP-P, the mean BCI-Utility reaches its maximum at the 6th sequence, when the model already attains a high mean accuracy, and gradually decreases thereafter, even though mean accuracy continues to rise until the 15th sequence. This illustrates an important trade-off. The optimal operating point for a BCI system is not necessarily the point with the highest accuracy, but the point that delivers the best balance between speed and accuracy \citep{ma2023bci}. The proposed method performs strongly on both metrics, achieving the highest accuracy while also maximizing user-centric BCI-Utility.

\subsection{Predictive Utility and Heterogeneity of Channel Interactions}

\subsubsection{When do channel interactions improve prediction performance?}
We investigate which group of participants are more likely to benefit from incorporating the signal interaction (SI) into the prediction model. Using questionnaire data collected in the same study by \citet{won2022eeg}, we compare the median accuracy achieved by SIRTGP-P and RTGP-P, stratified by participants' responses. Participants with prior exposure to BCI or similar biofeedback experiments show a 5\% improvement in character-level accuracy when SI is included, compared to models without SI. Similarly, participants who reported abstaining from alcohol 24 hours prior to the experiment achieved a 7\% gain with SI, with an maximum of 14\% improvement. Additional factors associated with an increased benefit from modeling SI include being in good mental state (+4\%) and good eye condition (+4\%) after testing phase, finding it hard to concentrate (+5\%), and perceiving the BCI task as difficult (+4\%). 

These findings align with the existing neuroscience literature, which has reported increased brain connectivity or regional interactions when performing familiar tasks \citep{terstege2022brain, noad2024familiarity}, being in good mental and physical state \citep{douw2014healthy, ismaylova2018associations, costumero2020opening}, making cognitive effort during tasks \citep{aben2020cognitive, wang2024changes}, and the disruptive effects of alcohol consumption on brain connectivity \citep{shokri2017alcohol, elton2021risk}. Our results suggest that the benefit of incorporating SI in prediction models may vary between participants. Future adaptive BCI experiments could consider tailoring the use of SI predictors based on previous  participant experience, such as alcohol consumption and reported mental states.

\begin{figure}[t]
    \spacingset{1}
    \centering
    \includegraphics[width=\linewidth]{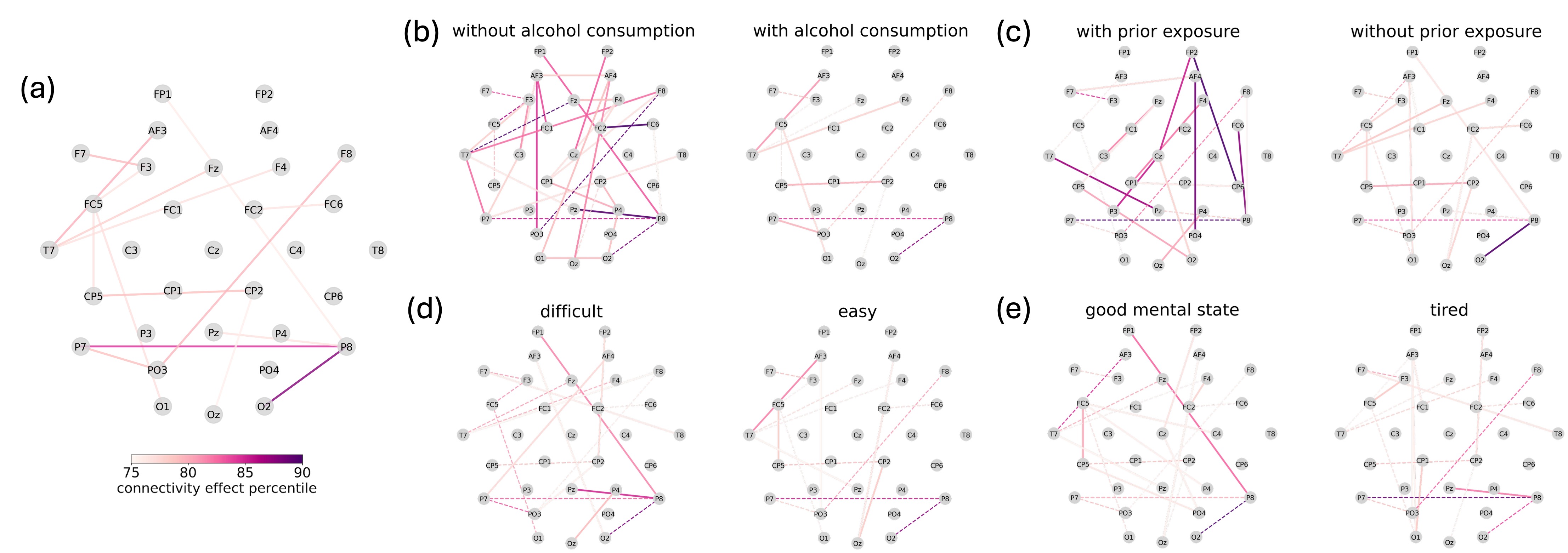}
    \caption{\highlight{Average percentiles for channel interaction selection probability: (a) over all 55 participants, (b) over participants without/with alcohol consumption in 24 hours, (c) over participants with/without prior exposure to similar experiments, (d) over participants who reported the BCI was difficult/easy after experiment, and (e) over participants who reported was in good mental state/tired. For clarity, only edges with percentiles greater than 75\% are shown. In (b)-(e), dashed lines show common selection by both subgroups, and solid lines show the unique solution to the subgroup.}}
    \label{fig:connectivity_percentile}
\end{figure}

\subsubsection{Heterogeneity in channel interaction effects} Based on the estimated SIRTGP-P signal interaction selection between-channels, we examine which pair of channels consistently contributed to the prediction performance in the 55 participants. Because a separate model is fitted for each individual, the posterior selection probability of channel interaction may not be directly comparable across individuals, we compute the percentile of the posterior channel interaction selection probability for each individual and then average these percentiles across participants for each channel pair. \highlight{Figure \ref{fig:connectivity_percentile}(a) displays the channel pairs whose average percentile exceeded 75\% over all 55 participants.}

\highlight{We observe that the interaction between P8 and O2 has one of the highest percentiles of selection probability, as does the interaction between P7 and P8. These patterns are consistent with existing neuroscience findings and are particularly relevant for the P300 BCI speller paradigm, which requires both visual-language processing and sustained attention. P7 and P8 are located at the temporal-parietal border, with P7 over the left hemisphere and P8 over the right hemisphere; O2 is located over the occipital region \citep{phang2020intralobular,kurihara2022relationship}. The P8-O2 pair reflects parieto-occipital connectivity, and the strong effect observed here is consistent with prior evidence of parieto-occipital coupling during stimulus-driven visuospatial attention and during tasks involving visual and language related processing \citep{indovina2004occipital, aguilar2025spatial}. The P7-P8 pair reflects inter-hemispheric connectivity between left and right temporo-parietal regions, areas implicated in visuospatial attention and sensorimotor integration \citep{freedman2018integrative}; it also reflects language-related processing given the proximity of these sites to posterior temporal language areas, particularly on the left \citep{michael2001fmri, trimmel2018left}.}

To examine how channel interaction effects vary between participant groups, we plotted the average percentiles of probability of interaction selection stratified by questionnaire responses \highlight{in Figure \ref{fig:connectivity_percentile}(b)-(e)}. The plotted network represents the subset of connectivity effects that are consistently identified as predictive. A denser network indicates that predictive information is distributed across a broader set of channel interactions, suggesting a widespread and active network-level engagement during the task. Conversely, a sparser network would imply that only a small number of specific channel-pair interactions are reliably informative. Participants who abstained from alcohol 24 hours prior to the experiment, had prior experience with similar tasks, reported the BCI typing task as difficult, or felt relaxed during the experiment exhibit denser networks on average than their respective counterparts. This suggests that channel interactions are more predictive and informative in these groups, consistent with prior evidence linking increased brain connectivity to task familiarity, mental and physical state, cognitive effort, and the absence of alcohol-related disruptions \citep{aben2020cognitive, elton2021risk, noad2024familiarity, wang2024changes}.

\highlight{The P7-P8 pair identified in Figure \ref{fig:connectivity_percentile}(a) appears in all stratified plots, while the P8-O2 pair is observed in all stratified plots except for participants with prior exposure. Nevertheless, the relative strength of P7-P8 and P8-O2 vary considerably across groups, despite their presence in most plots.} In general, the stratified plots exhibit substantially different patterns compared to the aggregate plot. This heterogeneity of signal interactions across participant groups highlights the value of the proposed method and helps explain why models that account for signal interactions can outperform those that do not.

\begin{figure}[t!]
    \spacingset{1}
    \centering
    \includegraphics[width=0.9\linewidth]{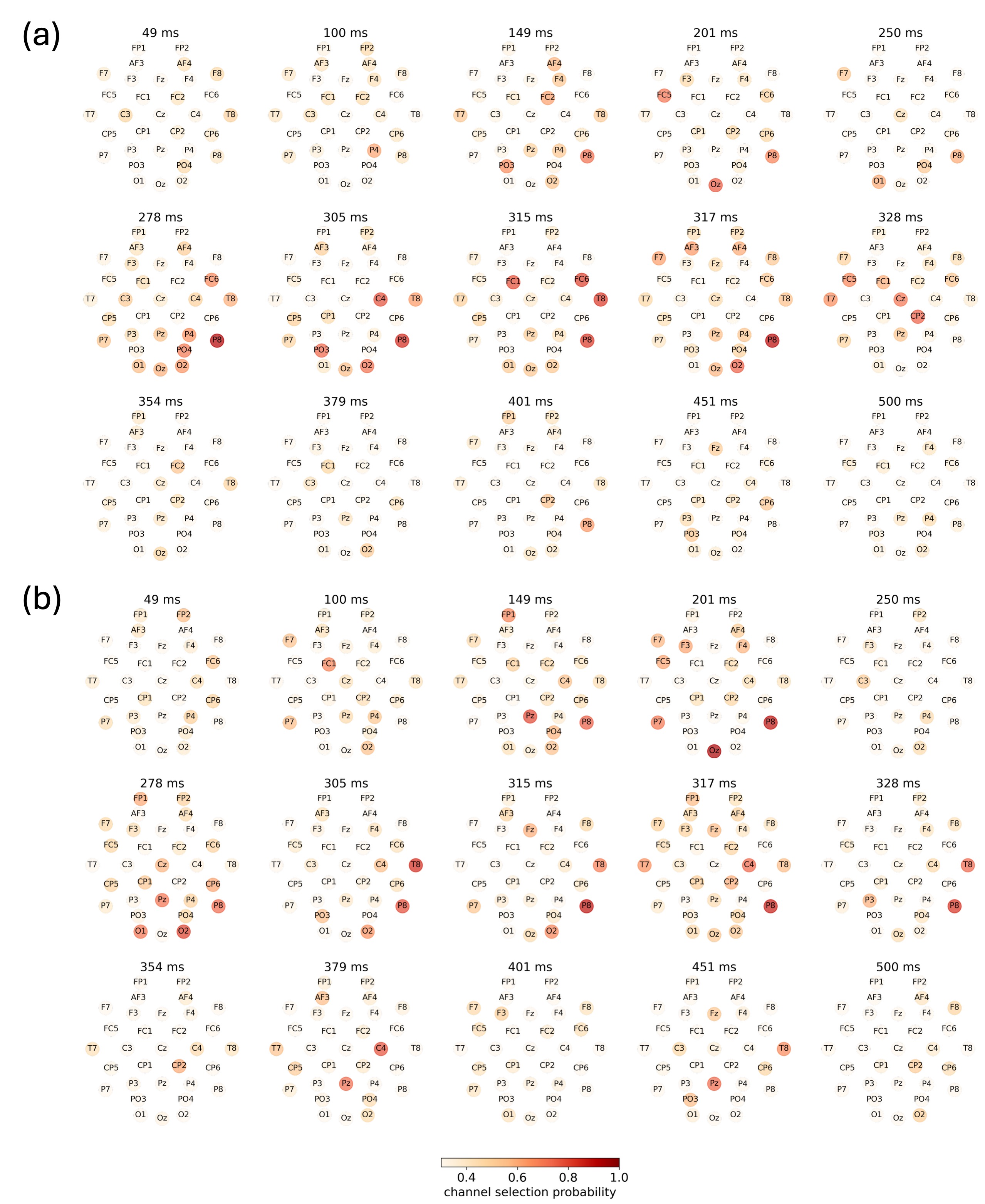}
    \caption{Posterior channel selection probability at different time points after the onset of stimulus presentation for one subject based on (a) SIRTGP-P and (b) RTGP-P. }
    \label{fig:channel_selection}
\end{figure}

\begin{figure}[t]
    \spacingset{1}
    \centering
    \includegraphics[width=\linewidth]{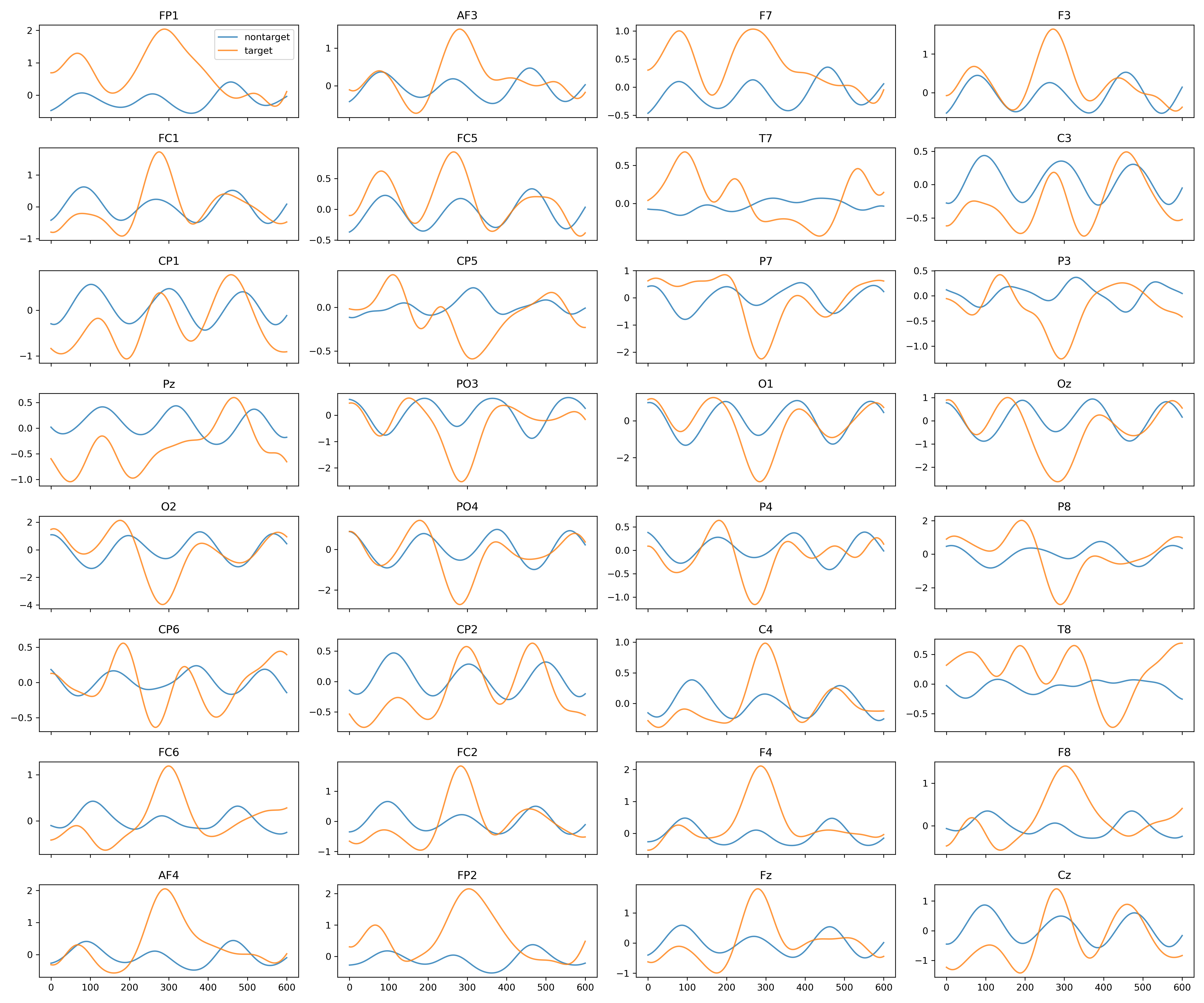}
    \caption{Average target (orange) and nontarget (blue) signals for all 32 channels from the same subject of Figure \ref{fig:channel_selection}. }
    \label{fig:signal}
\end{figure}

\subsection{Temporal dynamics of channel selection} 
To illustrate how the proposed method identifies informative channels over time, we present the channel selection results for a representative subject with prior exposure to BCI tasks. We focus on this subject because our group-level analysis indicates that participants with prior BCI exposure tend to benefit more from incorporating signal interaction (SI), making this subject useful for illustrating how SI influences channel selection over time. For this subject, SIRTGP-P improves prediction accuracy from 71\% to 93\%, relative to RTGP-P without SI. Figure \ref{fig:channel_selection} shows the posterior channel selection probabilities at multiple time points following stimulus onset, with panel (a) displaying results from SIRTGP-P with SI and panel (b) showing the corresponding results from RTGP-P without SI. The observed improvement is consistent with the overall pattern among participants with prior BCI exposure and allows us to examine the temporal and spatial channel-selection patterns underlying the benefit of incorporating SI.

Figure \ref{fig:signal} presents the average target and nontarget signals across for 32 channels for the same subject. The target responses exhibit a clear and typical P300 morphology at approximately 300 ms post-stimulus (spanning 200-400 ms), which serves as the primary discriminative feature between target and nontarget flashes. A notable pattern emerges when comparing the two panels in Figure \ref{fig:channel_selection}. The SI-enabled model in panel (a) identifies a larger set of informative channels around the P300 latency window. In contrast, panel (b) shows broadly similar, but consistently weaker selection patterns when SI is omitted around 300 ms. These results provide a concrete example in which explicitly modeling signal interaction enables the method to detect more spatially distributed task-relevant features and to select channels that are temporally consistent, which in turn, improves predictive performance.

\begin{revision}
To investigate which channels are most consistently selected across 55 participants, we compute the average percentile of posterior channel selection probability over participants. Figure \ref{fig:avg_channel_sel_prob} visualizes these probabilities at different time points. The channels with the highest percentiles are P8, O1, PO4, Oz, Pz, CP2, and O2. These channels correspond to parietal and occipital regions, both part of the posterior cortical cluster, which is consistent with the neuroscience literature \citep{indovina2004occipital, aguilar2025spatial}. Moreover, this finding aligns with prior BCI research \citep{ma2022bayesian, matianwen2026bayesian} identifying PO7, PO8, and Oz as important channels using other datasets. Although PO7 and PO8 are absent from this dataset's channel list, they are spatially proximate to O1, O2, PO4, and P8.
\begin{figure}[t]
    \centering
    \spacingset{1}
    \includegraphics[width=0.9\linewidth]{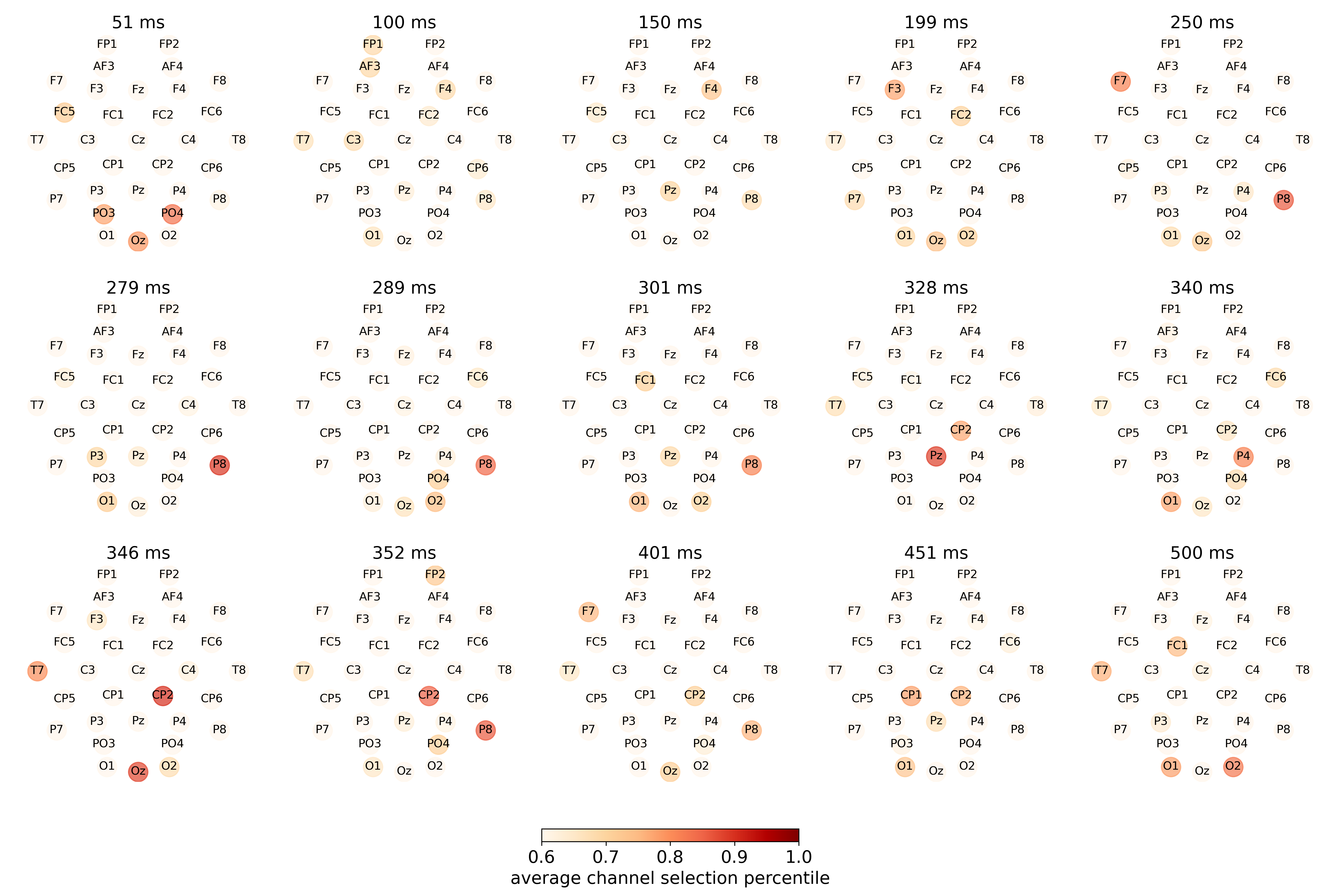}
    \caption{\highlight{Average percentile of posterior channel selection probability over 55 participants at different time points after the onset of stimulus presentation for each channel.}}
    \label{fig:avg_channel_sel_prob}
\end{figure}
\end{revision}

\begin{revision}
\section{Discussions}
In this study, we propose a Bayesian time-varying regression model with signal interactions via the relaxed-thresholded Gaussian process (SIRTGP) priors for P300 BCI speller. The proposed RTGP prior provides a unified framework that can recover the behaviors of a GP, a soft-thresholded GP, and a thresholded GP with proper relaxing parameters. The application of SIRTGP on a publicly available dataset demonstrates its real-world advantage compared to common predictive models used in P300 speller. Furthermore, through the proposed framework, we identify key channels and channel pairs that contribute to P300 detection, which offers insights for future BCI study and neural signal modeling.

The empirical results confirm that incorporating signal interactions yields meaningful gains in BCI performance. Comparing the SIRTGP model to the RTGP model, which shares the same architecture but excludes channel interactions, we observe an improvement on character-level prediction accuracy. This gain is further reflected in throughput. SIRTGP-P achieves the highest median BCI-Utility across all 15 sequences and reaches its peak throughput at the 6th sequence repetition, demonstrating that the benefit of modeling channel interactions extends beyond raw accuracy to the user-centric efficiency of the system.

However, the benefit of incorporating signal interactions is not uniform across participants. Certain subgroups benefit considerably more than others. For example, participants who abstained from alcohol in the 24 hours preceding the experiment showed up to an 14\% improvement in accuracy when signal interactions were included. These patterns align with the neuroscience literature linking increased brain network connectivity to task familiarity, favorable mental state, sustained cognitive effort, and the absence of alcohol-related disruptions to functional connectivity. The stratified channel interaction networks in Figure \ref{fig:connectivity_percentile}(b)-(e) further illustrate this heterogeneity. Subgroups with larger accuracy gains tend to exhibit denser and more widespread connectivity effect patterns, indicating that inter-channel interactions are more informative and more consistently predictive in these populations.

These findings have direct implications for BCI system design. The observed heterogeneity in the value of signal interaction modeling suggests that a one-size-fits-all approach may not be optimal. Instead, the decision of whether and how to incorporate channel interactions could be tailored to individual users based on readily available contextual information, such as self-reported physical and mental states, alcohol consumption, or prior BCI experience. This opens the door to adaptive and personalized BCI systems that dynamically adjust their model structure to the characteristics of each user, with the potential to improve both accuracy and throughput in real-world deployment.

An important direction for future work is to move beyond pairwise channel interactions and consider higher-order interaction structures among EEG channels. While the current framework captures dependencies between pairs of channels, neural information processing is believed to involve coordinated activity across distributed brain networks. Existing studies have demonstrated the importance of higher-order interactions in neural systems beyond pairwise relationships \citep{yu2011higher}. One promising direction is to represent such higher-order dependencies using hypergraph-based frameworks \citep{yu2025hypergraph}, allowing interactions among multiple channels to be modeled directly rather than through pairwise approximations. However, extending interaction modeling beyond pairs introduces substantial statistical and computational challenges, as the interaction space grows rapidly with the interaction order. Developing scalable methods for identifying sparse higher-order interaction patterns remains an important topic for future research.

The proposed framework is developed and validated specifically for EEG-based P300 speller BCIs. However, its application is not tied to EEG and may generalize to general neuroimaging applications. The methodological contribution, relaxed-thresholded Gaussian process (RTGP) prior, is a general prior that provides both flexibility and sparsity. As formalized in Proposition 1, RTGP can adapt to the underlying signal structure in a data-driven manner while retaining a conjugate and closed-form full conditional that supports efficient Gibbs sampling even in high-dimensional settings. These properties make RTGP applicable to a broad class of problem involving functional or spatially indexed predictors where sparsity is plausible but not guaranteed to be sharp, and where spatial and/or temporal dependence structures need to be accounted for. 
\end{revision}

\section*{Data and Code Availability}

The data used this study is from a publicly available EEG dataset for RSVP and P300 speller brain–computer interface experiments \citep{won2022eeg}. The dataset is openly accessible at https://doi.org/10.6084/m9.figshare.c.5769449. Our method has a Python implementation and is available at https://anonymous.4open.science/r/SIRTGP-6402. 

\spacingset{1.5}

\bibliographystyle{agsm} 
\bibliography{reference}       

\end{document}